\documentclass[letterpaper, 10 pt, conference]{ieeeconf}  % Comment this line out if you need a4paper
\pdfoutput=1

\IEEEoverridecommandlockouts                              % This command is only needed if
\usepackage{amsmath, amssymb}
\usepackage[utf8]{inputenc}
\usepackage[pdftex]{graphicx}
\graphicspath{{./Figures/}}
\usepackage{epstopdf}
\usepackage{float}
\usepackage{booktabs}
\usepackage{tabularx}
\usepackage{subcaption} % For subfigures
\usepackage{amsthm} % For theorems, assumptions, remarks, etc
    
    \newtheorem{assumption}{Assumption}

    \newtheorem{remark}{Remark}

\usepackage[ruled, lined, ,longend, linesnumbered]{algorithm2e}
\SetKwInOut{Input}{Input}
\SetKwInOut{Declares}{Declares}
\SetKwInOut{Require}{Require}
\SetKw{KwBy}{by}

\newcommand{\cc}[1]{{\mathcal{#1}}} % Short for giving it the font for sets
\def\R{ {\rm \,I\!R} } % Set of real numbers
\def\Z{{\mathbb{Z}}} % Set of integers
\def\inR#1{\in\R^{#1}} % Vector space of real numbers of dimension #1
\def\vv#1{{ \rm \bf{#1}}} % Bold
\newcommand{\T}{^\top} % Transpose
\def\sp#1#2{\langle #1,#2\rangle } % Scalar product
\def\Sum#1#2{\sum\limits_{#1}^{#2}} % Sum with limits
\DeclareMathSymbol{\shortminus}{\mathbin}{AMSa}{"39} % Short minus sign
\def\bmat#1{\left[\begin{array}{#1}} % Begin matrix
\def\emat{\end{array}\right]} % End matrix
\newcommand {\bsis} {\left\{ \begin{array} }
\newcommand {\esis} {\end{array}\right.}
\newcommand{\bRlist}{\renewcommand{\labelenumi}{(\roman{enumi})} \begin{enumerate}} % Begin enumerate with roman numbers
\newcommand{\eRlist}{\end{enumerate} \renewcommand{\labelenumi}{\arabic{enumi}}} % End enumerate with roman numbers
\def\tx{\tilde{x}}
\def\tu{\tilde{u}}

\def\figwidth{0.46}

\usepackage{xcolor}
\title{Real-time implementation of MPC for tracking in embedded systems: Application to a two-wheeled inverted pendulum}
\author{Pablo Krupa, Jose Camara, Ignacio Alvarado, Daniel Limon, Teodoro Alamo%
\thanks{Systems Engineering and Automation department, University of Seville, Spain. The corresponding author is Pablo Krupa (\texttt{pkrupa@us.es}).}%
\thanks{This work was supported in part by the Agencia Estatal de Investigación (AEI) under Grant PID2019-106212RB-C41/AEI/10.13039/501100011033, by MINERCO-Spain and FEDER funds under Grant DPI2016-76493-C3-1-R and by the MCIU-Spain and FSE under Grant FPI-2017.}%
}

\begin{document}
% Fakesection Maketitle
\maketitle

% Fakesection Abstract
\begin{abstract}
    This article presents the real-time implementation of the model predictive control for tracking formulation to control a two-wheeled inverted pendulum robot.
    This formulation offers several advantages over standard MPC formulations at the expense of the addition of a small number of decision variables, which complicates the inner structure of the matrices of the optimization problem.
    We implement a sparse solver, based on an extension of the alternating direction method of multipliers, in the system's embedded hardware.
    The results indicate that the solver is suitable for controlling a real system with sample times in the range of milliseconds using current, readily-available hardware.
\end{abstract}

\subsubsection*{Keywords}
Model predictive control, embedded system, extended ADMM

\section{Introduction} \label{sec:introduction}

The implementation of model predictive control (MPC) in embedded systems has been a widely researched topic in recent years due to the interest of being able to use this advanced control strategy to control real systems using the currently available embedded hardware. One of the main challenges that needs to be overcome is the fact that MPC requires solving an optimization problem at each sample time, which can become an issue in systems with fast dynamics, especially when considering the low computational and memory resources typically available in embedded systems.

Recently, significant advances have been made in this field thanks to the development of optimization algorithms suitable for their implementation in embedded systems. Some examples of these tools being used to implement MPC in embedded systems include \cite{Hartley_IETCST_2014, Huyck_MED_2012, Krupa_IFAC_20}. Additionally, other authors propose algorithms that are particularly tailored to the MPC optimization problem, such as in \cite{Krupa_TCST_20,Lucia_IETII_2018,Shukla_SD_2017,Park_IPA_SQP_2014}. Finally, another approach is to use \textit{explicit} MPC \cite{Tondel_A_2003}, which computes the solution of the parametric MPC optimization problem offline and stores it online as a lookup table. However, this is only suitable for systems with few states and a moderate number of constraints.

A common theme among the current research on this topic, which is shared by the previously cited papers, is that they typically only consider standard MPC formulations. This paper, on the other hand, presents an implementation of a non-standard MPC formulation called MPC \textit{for tracking} (MPCT)~\cite{Ferramosca_A_2009}, which adds an \textit{artificial reference} as an additional decision variable of the optimization problem. This formulation provides a series of advantages that make its implementation in embedded systems particularly interesting.

Firstly, a common issue of standard MPC formulations with stability guarantees is that the domain of attraction of the controller can become insufficient if the prediction horizon is chosen too small. However, the use of small prediction horizons is desirable in order to help overcome the computational and memory limitations typically imposed by embedded systems. The MPCT formulation provides significantly larger domains of attraction than standard MPC formulations \cite{Limon_A_2008}, especially for small prediction horizons.

Secondly, it intrinsically deals with references that are not attainable (i.e. that are not a steady state of the system or that violate the system constraints) \cite{Limon_A_2008}. In this case, it will steer the system to the ``closest" attainable steady state to the reference, where the ``closeness" is determined by the selection of its cost function matrices. Additionally, it also guarantees recursive feasibility of the closed-loop system even in the event of a sudden reference change \cite{Ferramosca_A_2009}.

However, these advantages come at the cost of the addition of the \textit{artificial reference} as new decision variables, which complicates the inner structure of the matrices of the quadratic programming problem when compared to the standard MPC formulations.

In \cite{Krupa_arxiv_MPCT_20} the authors presented a sparse solver for the MPCT formulation based on an extension of the classical alternating direction method of multipliers (ADMM) \cite{Boyd_FTML_2011} to problems with three separable functions in the objective function \cite{Cai_EADMM_2017}. 
The use of this method resulted in the ingredients of the algorithm having simple structures that could be exploited using a similar approach to the one used in \cite{Krupa_TCST_20}, which presented sparse solvers for standard MPC formulations.
This lead to a sparse solver with a small iteration complexity and a small memory footprint that was included in the Spcies toolbox \cite{Spcies} for Matlab, which is available at \url{https://github.com/GepocUS/Spcies}.

This paper presents the implementation of the above MPCT solver in a Raspberry Pi to control a two-wheeled inverted pendulum robot with a sample time of $20$ milliseconds.
The closed-loop results suggest that the solver is suitable for its implementation in current, readily-available hardware for controlling systems with fast dynamics.

This paper is organized as follows.
Section \ref{sec:problem:formulation} provides the problem formulation.
The MPCT formulation is described in Section \ref{sec:MPCT}.
For completeness, a brief description of the solver is presented in Section \ref{sec:solver}.
The two-wheeled inverted pendulum robot and the closed-loop results are shown in Section \ref{sec:case:study}.
Finally, conclusions are provided in Section \ref{sec:conclusions}.

% Fakesection Notation
\subsubsection*{Notation}
Given two integers $i$ and $j$ with ${j \geq i}$, $\Z_i^j$ denotes the set of integer numbers from $i$ to $j$, i.e. ${\Z_i^j \doteq \{i, i+1, \dots, j-1, j\}}$.
Given two vectors $x \inR{n}$ and $y \inR{n}$, $x \leq (\geq) \; y$ denotes componentwise inequalities; and $\sp{x}{y}$ denotes their standard inner product.
% Given a matrix $A \inR{n \times m}$, $A_{i,j}$ denotes its element $(i, j)$, $A\T$ its transposed and $A^{-1}$ its inverse (if it is non-singular).
For a vector $x\inR{n}$ and a positive definite matrix $A \inR{n \times n}$, $\|x\| \doteq \sqrt{\sp{x}{x}}$, $\|x\|_A$ is its weighted Euclidean norm $\|x\|_A \doteq \sqrt{\sp{x}{A x}}$, and $\| x \|_\infty \doteq \max_{i = 1 \dots n}{| x_{(i)} |}$, where $x_{(i)}$ is the $i$-th element of $x$, is its $\ell_\infty$-norm.
We denote by $(x_{1}, x_{2}, \dots, x_{N})$ the column vector formed by the concatenation of column vectors $x_{1}$ to $x_{N}$.
Given scalars and/or matrices $M_1, M_2, \dots, M_N$, we denote by $\texttt{diag}(M_1, M_2, \dots, M_N)$ the block diagonal matrix formed by the diagonal concatenation of $M_1$ to $M_N$.

\section{Problem Formulation} \label{sec:problem:formulation}

We consider a system described by a linear time-invariant state-space model
\begin{equation} \label{eq:model}
    x_{k+1} = A x_k + B u_k,
\end{equation}
where $x_k \inR{n}$ and $u_k \inR{m}$ are the state and input of the system at sample time $k$, respectively, that are subject to the box constraints
\begin{equation} \label{eq:constraints}
    \underline{x} \leq x_k \leq \overline{x}, \quad \underline{u} \leq u_k \leq \overline{u}.
\end{equation}

The control objective is to steer the system to the reference $(x_r, u_r)$ given by the user. This will only be possible if the reference is a steady state of the system that satisfies the constraints \eqref{eq:constraints}, i.e., if it is an \textit{admissible} steady state. Otherwise, we wish to steer the system to the closest admissible steady state to $(x_r, u_r)$, for a certain criterion of closeness.

\section{Model predictive control for tracking} \label{sec:MPCT}

This section describes the particular MPC formulation that we consider in this paper, which is called \textit{MPC for tracking}~\cite{Ferramosca_A_2009}.
For a given control horizon $N$, a current state $x \inR{n}$ and a reference $(x_r, u_r)$, the control law of the MPCT controller is derived from the solution of the optimization problem
\begin{subequations} \label{eq:MPCT} % MPCT
\begin{align}  
    \min\limits_{\substack{\vv{x}, \vv{u},\\ x_s, u_s}} \;& \Sum{i = 0}{N-1} \| x_i {-} x_s \|^2_Q {+} \| u_i {-} u_s \|^2_R {+} \| x_s {-} x_r \|^2_T {+} \| u_s {-} u_r \|^2_S \\
    s.t.& \; x_0 = x \label{eq:MPCT:initial} \\
        & \; x_{i+1} = A x_i + B u_i, \; i\in\Z_0^{N-1} \label{eq:MPCT:prediction} \\
        & \; \underline{x} \leq x_i \leq \overline{x}, \; i\in\Z_1^{N-1} \\
        & \; \underline{u} \leq u_i \leq \overline{u}, \; i\in\Z_0^{N-1} \\
        & \; x_s = A x_s + B u_s \label{eq:MPCT:steady:state}\\
        & \; \underline{x} + \varepsilon_x \leq x_s \leq \overline{x} - \varepsilon_x \label{eq:MPCT:ineq:x_s} \\
        & \; \underline{u} + \varepsilon_u \leq u_s \leq \overline{u} - \varepsilon_u \label{eq:MPCT:ineq:u_s} \\
        & \; x_N = x_s, \label{eq:MPCT:terminal}
\end{align}
\end{subequations}
where $\vv{x} = (x_0, x_1, \dots, x_{N-1})$ and $\vv{u} = (u_0, u_1, \dots, u_{N-1})$ are the predicted states and control actions throughout the prediction horizon, respectively; $(x_s, u_s)$ is the artificial reference; $\varepsilon_x \inR{n}$ and $\varepsilon_u \inR{m}$ are vectors with arbitrarily small positive components which are added to avoid a (possible) loss of controllability in the event of active constraints at the equilibrium point \cite{Limon_A_2008}; and the positive definite matrices $Q$, $R$, $T$ and $S$ are the cost function~matrices.

The main difference between MPCT and standard MPC formulations is the introduction of the artificial reference $(x_s, u_s)$ as additional decision variables. As can be seen in~\eqref{eq:MPCT}, the discrepancy between the predicted states and control actions with the artificial reference is penalized with matrices $Q$ and $R$. Additionally, the discrepancy between the artificial reference and the reference $(x_r, u_r)$ given by the user is penalized with matrices $T$ and $S$.

%As mentioned in the introduction, this formulation has several advantages when compared to standard MPC formulations, such as its guaranteed recursive feasibility even in the event of a sudden change of the reference $(x_r, u_r)$, or the fact that it provides significant improvements of the domain of attraction, especially for small prediction horizons. Additionally, it steers the closed-loop system to the admissible steady state $(x_{ss}, u_{ss})$ that minimizes the distance $\| x_{ss} - x_r \|^2_T + \| u_{ss} - u_r \|^2_S$  \cite{Ferramosca_A_2009, Limon_A_2008}.

\section{Embedded solver for MPCT} \label{sec:solver}

This section briefly presents the sparse solver implemented in the embedded system, which is a particularization of the \textit{extended ADMM} algorithm \cite{Cai_EADMM_2017} to the optimization problem \eqref{eq:MPCT}.
This solver, which was originally presented in \cite{Krupa_arxiv_MPCT_20} and is available at \cite{Spcies}, exploits the structure of the problem to attain a very small memory footprint and an efficient implementation.
Due to space considerations, and to not repeat the results presented in \cite{Krupa_arxiv_MPCT_20}, only a very brief outline of the solver is presented here.
We refer the reader to the above reference for an in-depth explanation.

\subsection{Extended ADMM} \label{sec:solver:EADMM}

The extended ADMM algorithm is, as its name suggests, an extension of the classical ADMM algorithm \cite{Boyd_FTML_2011} to problems with more than two separable functions in the objective function. Specifically, we show its application to objective functions that are the sum o three separable functions \cite{Cai_EADMM_2017}.

Let $\theta_i : \R^{n_i} \rightarrow \R$ for $i \in\Z_1^3$ be convex functions; $\cc{Z}_i \subseteq \R^{n_i}$ for $i \in\Z_1^3$ be closed convex sets; $C_i \inR{m_z \times n_i}$ for $i \in\Z_1^3$; and $b \inR{m_z}$. Consider the optimization problem
\begin{subequations} \label{eq:EADMM:optimization:problem}
\begin{align}
    \min\limits_{z_1, z_2, z_3} &\Sum{i = 1}{3} \theta_i(z_i) \\
    s.t.& \; \Sum{i = 1}{3} C_i z_i = b \\
        & \; z_i \in \cc{Z}_i, \; i\in\Z_1^3,
\end{align}
\end{subequations}
where $z_i\inR{n_i}$ for $i \in\Z_1^3$ are the decision variables, and let its augmented Lagrangian $\cc{L}_\rho(z_1, z_2, z_3, \lambda)$ be given by
\begin{equation*}
    \cc{L}_\rho(\cdot) = \Sum{i = 1}{3} \theta_i(z_i) {+} \left\langle \lambda, \Sum{i = 1}{3} C_i z_i {-} b \right\rangle {+} \frac{\rho}{2} \left\| \, \Sum{i = 1}{3} C_i z_i {-} b \, \right\|^2,
\end{equation*}
where $\lambda \inR{m_z}$ are the dual variables and the scalar $\rho > 0$ is the penalty parameter.

Algorithm \ref{alg:EADMM} shows the implementation of the extended ADMM algorithm. It returns a suboptimal solution $(\tilde z_1^*,\tilde z_2^*,\tilde z_3^*)$ of  problem \eqref{eq:EADMM:optimization:problem} (assuming a solution point exists) as well as a suboptimal dual variable $\tilde \lambda^*$, where the suboptimality is determined by the exit tolerance $\epsilon > 0$, since the exit conditions of step \ref{alg:EADMM:step:exit} serve as a measure of the optimality of the current iterate \cite[\S 5]{Cai_EADMM_2017}. The superscript $k$ is used to indicate the value of the variable at iteration $k$ of the algorithm.

\begin{algorithm}[t]
    \DontPrintSemicolon
    \caption{Extended ADMM} \label{alg:EADMM}
    \Require{$z_2^0$, $z_3^0$, $\lambda^0$, $\rho > 0$, $\epsilon > 0$ \label{alg:EADMM:step:initial}}
    $k \gets 0$\;
    \Repeat{$\| \Gamma \|_\infty {\leq} \epsilon, \|  z_2^{k} {-} z_2^{k-1} \|_\infty {\leq} \epsilon, \| z_3^{k} {-} z_3^{k-1} \|_\infty {\leq} \epsilon$ \label{alg:EADMM:step:exit}}{
        $z_1^{k+1} {\gets} \arg\min\limits_{z_1} \{ \cc{L}_\rho(z_1, z_2^k, z_3^k, \lambda^k) \, | \, z_1 {\in} \cc{Z}_1 \}$\; \label{alg:EADMM:step:z_1}
        $z_2^{k+1} {\gets} \arg\min\limits_{z_2} \{ \cc{L}_\rho(z_1^{k+1}, z_2, z_3^k, \lambda^k) \, | \, z_2 {\in} \cc{Z}_2 \}$\; \label{alg:EADMM:step:z_2}
        $z_3^{k+1} {\gets} \arg\min\limits_{z_3} \{ \cc{L}_\rho(z_1^{k+1}, z_2^{k+1}, z_3, \lambda^k) \, | \, z_3 {\in} \cc{Z}_3 \}$\; \label{alg:EADMM:step:z_3}
        $\Gamma \gets \Sum{i = 1}{3} C_i z_i^{k+1} - b$\; \label{alg:EADMM:step:residual}
        $\lambda^{k+1} \gets \lambda^k + \rho \Gamma $\; \label{alg:EADMM:step:lambda}
        $k \gets k + 1$\;
    }
    \KwOut{$\tilde z_1^* {\gets} z_1^{k}$, $\tilde z_2^* {\gets} z_2^{k}$, $\tilde z_3^* {\gets} z_3^{k}$, $\tilde \lambda^* {\gets} \lambda^{k}$}
\end{algorithm}

The extended ADMM does not necessarily converge under the same assumptions as standard ADMM, as shown in \cite{Chen_EADMM_convergence_2016}. In order to prove its convergence, additional conditions are required. In particular, in \cite[Theorem 3.1]{Cai_EADMM_2017} it was shown that the extended ADMM algorithm applied to \eqref{eq:EADMM:optimization:problem} converges under the following assumption if $\rho$ is chosen appropriately (as stated in the cited theorem).

\begin{assumption}[\cite{Cai_EADMM_2017}, Assumption 3.1] \label{ass:EADMM}
    The functions $\theta_1$ and $\theta_2$ are convex; function $\theta_3$ is strongly convex; and $C_1$ and $C_2$ are full column rank.
\end{assumption}

\subsection{Solving MPCT using EADMM} \label{sec:solver:embedded}

This section explains how problem \eqref{eq:MPCT} can be recast into \eqref{eq:EADMM:optimization:problem} by a proper selection of decision variables.
We do so by defining variables $\tx_i \doteq x_i - x_s$ and $\tu_i \doteq u_i - u_s$, which lets us rewrite \eqref{eq:MPCT} as:
\begin{subequations} \label{eq:MPCT:transformed} % MPCT
\begin{align}  
    \min\limits_{\substack{ \tilde{\vv{x}}, \tilde{\vv{u}}, \vv{x},\\ \vv{u}, x_s, u_s} } \; &\Sum{i = 0}{N} \| \tx_i \|^2_Q +  \| \tu_i \|^2_R + \| x_s - x_r \|^2_T + \| u_s - u_r \|^2_S \\
    s.t.& \; x_0 = x \label{eq:MPCT:transformed:initial} \\
        & \; \tx_{i+1} = A \tx_i + B \tu_i, \; i\in\Z_0^{N-1} \label{eq:MPCT:transformed:model} \\
        & \; \underline{x} \leq x_i \leq \overline{x}, \; i\in\Z_1^{N-1} \label{eq:MPCT:transformed:ineq:x} \\
        & \; \underline{u} \leq u_i \leq \overline{u}, \; i\in\Z_0^{N-1} \label{eq:MPCT:transformed:ineq:u}\\
        & \; \underline{x} + \varepsilon_x \leq x_N \leq \overline{x} - \varepsilon_x \label{eq:MPCT:transformed:ineq:x_N} \\
        & \; \underline{u} + \varepsilon_u \leq u_N \leq \overline{u} - \varepsilon_u \label{eq:MPCT:transformed:ineq:u_N}\\
        & \; x_s = A x_s + B u_s \label{eq:MPCT:transformed:xsus}\\
        & \; \tx_i + x_s - x_i = 0, \; i\in\Z_0^{N} \label{eq:MPCT:transformed:x}\\
        & \; \tu_i + u_s - u_i = 0, \; i\in\Z_0^{N} \label{eq:MPCT:transformed:u}\\
        & \; x_N = x_s \label{eq:MPCT:transformed:terminal:x_s} \\
        & \; u_N = u_s, \label{eq:MPCT:transformed:terminal:u_s}
\end{align}
\end{subequations}
where we are introducing the new decision variables ${\tilde{\vv{x}} = (\tx_0, \dots, \tx_N)}$ and ${\tilde{\vv{u}} = (\tu_0, \dots, \tu_N)}$. Note that we have extended the summations in the cost function (and some constraints) to $i = N$. This is necessary to be able to construct matrices $C_i$ of \eqref{eq:EADMM:optimization:problem} with a simple structure. However, note that this additional term does not affect the optimization problem due to constraints \eqref{eq:MPCT:transformed:terminal:x_s} and \eqref{eq:MPCT:transformed:terminal:u_s}.

Problem \eqref{eq:MPCT:transformed} can then be recast as \eqref{eq:EADMM:optimization:problem} by defining
\begin{subequations} \label{eq:MPCT:z:selection}
\begin{align}
    z_1 &= (x_0, u_0, x_1, u_1, \dots, x_{N-1}, u_{N-1}, x_N, u_N), \\
    z_2 &= (x_s, u_s), \\
    z_3 &= (\tx_0, \tu_0, \tx_1, \tu_1, \dots, \tx_{N-1}, \tu_{N-1}, \tx_N, \tu_N),
\end{align}
\end{subequations}
and casting the constraints \eqref{eq:MPCT:transformed:initial}, \eqref{eq:MPCT:transformed:x}, \eqref{eq:MPCT:transformed:u}, \eqref{eq:MPCT:transformed:terminal:x_s} and \eqref{eq:MPCT:transformed:terminal:u_s} into $C_i$ and $b$; constraints \eqref{eq:MPCT:transformed:ineq:x}-\eqref{eq:MPCT:transformed:ineq:u_N} into $\mathcal{Z}_1$, \eqref{eq:MPCT:transformed:xsus} into $\mathcal{Z}_2$ and \eqref{eq:MPCT:transformed:model} into $\mathcal{Z}_3$.
Then, the functions $\theta_i$ are given by $\theta_1(z_1) = 0$
\begin{align*}
    &\theta_2 (z_2) = \frac{1}{2} z_2\T \texttt{diag}(T, S) z_2 - (T x_r, S u_r)\T z_2, \\
    &\theta_3(z_3) = \frac{1}{2} z_3\T \texttt{diag}(Q, R, Q, R, \dots, Q, R) z_3,
\end{align*}

We note that our selection of $z_i$ and $C_i$ for $i \in\Z_1^3$ leads to a problem \eqref{eq:EADMM:optimization:problem} that satisfies Assumption \ref{ass:EADMM}.

The selection of the decision variables $z_1$, $z_2$ and $z_3$ as shown in \eqref{eq:MPCT:z:selection} leads to matrices $C_i$ with very simple structures (see \cite[Eq. (9)]{Krupa_arxiv_MPCT_20}) and to the optimization problems solved in steps \ref{alg:EADMM:step:z_1}, \ref{alg:EADMM:step:z_2} and \ref{alg:EADMM:step:z_3} of Algorithm \ref{alg:EADMM} to have explicit and computationally efficient solutions \cite[\S 5.2]{Krupa_arxiv_MPCT_20}.

\begin{remark}[\cite{Krupa_arxiv_MPCT_20}, Remark 1] \label{rem:rho:matrix}
In \cite[\S 5.2]{Stellato_OSQP_arXiv_2020} it was shown that the performance of ADMM can be improved if different values of $\rho$ are used to penalize some constraints more than others, i.e., by taking $\rho$ as a diagonal positive definite matrix.
In particular, for problem \eqref{eq:MPCT:transformed}, the convergence improves significantly if the equality constraints \eqref{eq:MPCT:transformed:initial}, \eqref{eq:MPCT:transformed:terminal:x_s}, \eqref{eq:MPCT:transformed:terminal:u_s}, \eqref{eq:MPCT:transformed:u} for $i = N$,
\eqref{eq:MPCT:transformed:x} for $i = 0$ and $i = N$, are penalized more than the others.
\end{remark}

\section{Case study} \label{sec:case:study}

\subsection{Two-wheeled inverted pendulum robot} \label{sec:case:study:system}

\begin{figure}[t]
    \centering
    \begin{minipage}{0.45\columnwidth}
    \includegraphics[width=0.6\columnwidth]{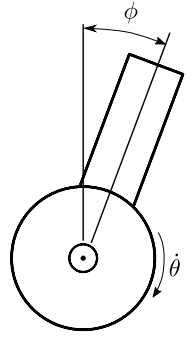}%
    \end{minipage}%
    \;
    \begin{minipage}{0.45\columnwidth}
    \includegraphics[width=0.9\columnwidth]{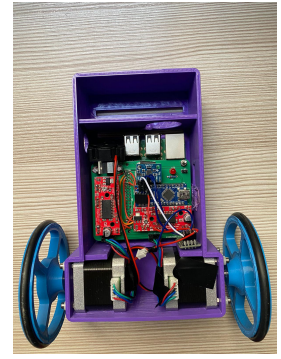}
    \end{minipage}
    \caption{Two-wheeled inverted pendulum robot. The angle $\phi$ is zero if the pendulum is in a vertical position.}
    \label{fig:segway}
\end{figure}

The two-wheeled inverted pendulum robot, which is shown in Figure~\ref{fig:segway}, is a two-wheeled vehicle based on the inverted pendulum configuration. The control objective is to control the horizontal speed of the vehicle whilst keeping it from toppling. Due to construction limitations of the robot, in this paper we only consider forward and backward velocities, i.e., both wheels have the same speed, making the robot incapable of rotating sideways.

The specifics of the robot, including its construction and components, are very similar to the description provided in \cite{Borja_IWC_20}. The chassis is made using a 3D printer, and its main components are: a Raspberry Pi, an inertial measuring unit MPU6050, an Arduino NANO, a microstepping motor driver A3967 and two step motors Nema 17. The main difference between this robot and the one described in \cite{Borja_IWC_20} is the inclusion of the Raspberry Pi for monitoring and controlling the system, as we explain in further detail in Section~\ref{sec:case:study:embedded}.

The non-linear dynamics of the systems, which are obtained by applying Lagrange's equation as in \cite[Appendix A]{Cecilia_2017}, are given by the ordinary differential equation  \cite[\S 6]{Borja_IWC_20}
\begin{equation} \label{eq:segway:ODE}
\begin{aligned}
    &(2a+c\cos{(\phi+\phi_{0})})\ddot{\theta}+(c\cos{(\phi+\phi_{0})}+2b)\ddot{\phi} \\ &-c\dot{\phi}^2\sin(\phi+\phi_{0})-d\sin(\phi+\phi_{0})=0,
\end{aligned}
\end{equation}
where $\phi$ is the tilt of the robot, $\theta$ is the angle of the wheels, $a=(3/2 m_r + 1/2)R^2$, $b=ML^2$, $c=RML$, $d=MgL$, $m_r$ is the mass of the wheels, $M$ is the mass of the robot without the wheels, $R$ is the wheel's radius, $L$ is the distance between the rotation axis of the wheels and the center of mass, $g = 9.81$m/s$^2$ is the gravitational acceleration and $\phi_0$ is the angle between the center of mass and the geometrical center. The robot used in this case study, which is shown in the right-hand-side of Figure \ref{fig:segway}, has the following values of the parameters: $m_r = 0.064$Kg, $M = 0.975$Kg, $R = 0.05$m, $L = 0.05$m and $\phi_0$ is unmeasured but known to be small.

The state of the system is given by $x = (\phi, \dot \phi, \dot \theta)$ and the control input is the angular acceleration of the wheels $u = \ddot \theta$. We consider the following constraints on the state and control input,
\begin{equation*} \label{eq:segway:constraints}
\begin{bmatrix}
-\frac{90}{360} 2\pi\\ 
-4\\ 
-60
\end{bmatrix}\leq \begin{bmatrix}
\phi\\ 
\dot{\phi}\\ 
\dot{\theta}
\end{bmatrix} \leq \begin{bmatrix}
\frac{90}{360} 2\pi\\ 
4\\ 
60
\end{bmatrix}, \; -80\leq \ddot{\theta} \leq 80,
\end{equation*}
where the units are given in radians and seconds, accordingly.

\subsection{Embedded system: Raspberry Pi} \label{sec:case:study:embedded}

The Rapsberry Pi is a low-cost embedded system that can be operated by a Linux-based operating system. In particular, we use the Raspbian operating system provided by the manufacturer, which is based on the Debian distribution. The model used in this case study is Raspberry Pi 3 Model B, which comes with a Quad Core 1.2GHz Broadcom BCM2837 64bit CPU.

The Raspberry Pi is used as the monitoring and control device of the robot. It receives the measurements of the tilt angle $\phi$ and the tilt angular speed $\dot \phi$ from the inertial measuring unit MPU6050, and the measurement of the speed of the wheels $\dot \theta$ from the Arduino NANO board. The control action is sent to the Arduino NANO board, which is in charge of applying the corresponding PWM signals to the step motors.

In order to ensure the real-time operation of the control system, we employ the Xenomai software, which is a dual-kernel configuration for Linux-based systems which considers the Linux kernel as an idle task, and that will ensure the accomplishment of the scheduled tasks within the given deadlines. In short, it provides a real-time framework to Linux-based systems, which we use to schedule the measurement and MPCT controller routines in real-time.

\subsection{Closed-loop results} \label{sec:case:study:results}

We design the MPCT controller, taking the parameters $N = 12$, $Q = 5 I_3$, $R = 1$, $T = 1000 I_3$ and $S = 5$. The exit tolerance of the EADMM algorithm is $\epsilon = 0.001$, and the penalty parameter is $\rho = 1000$ for the constraints listed in Remark \ref{rem:rho:matrix}, and $\rho = 5$ for the rest. The prediction model of the MPCT controller is obtained by linearizing the non-linear model \eqref{eq:segway:ODE} around the operating point $x^\circ = (0, 0, 0)$ and $u^\circ = (0)$, i.e., the stationary vertical position, for a sample time of $20$ms.

{\renewcommand{\arraystretch}{1.1}%
    \begin{table}[t]
    \centering
    \caption{Performance of EADMM}
    \label{tab:case:study:computation:performance}
    \setlength\tabcolsep{4pt}
    \setlength{\arrayrulewidth}{.1em}
    \setlength{\extrarowheight}{3pt}
    \begin{tabular}{c|cccccccccc}
    \multicolumn{1}{c}{} & \multicolumn{4}{c}{Iterations} & \multicolumn{4}{c}{Computation time (ms)} \\
    \cmidrule[1pt](lr){2-5} \cmidrule[1pt](lr){6-9} 
     \multicolumn{1}{c}{} & Max. & Min. & Med. & Avg. & Max. & Min. & Med. & Avg.\\
     \hline
     Fig. \ref{fig:pulse} & 44 & 1 & 15 & 15.12 & 8.64 & 0.195 & 2.99 & 2.99\\
     Fig. \ref{fig:ref} & 38 & 1 & 11 & 12.38 & 7.39 & 0.196 & 2.15 & 2.42 \\
     \hline
    \end{tabular}
\end{table}}

The solver is obtained from the Matlab toolbox \cite{Spcies}, which automatically generates code for solving different MPC formulations, including the solver discussed here.
The toolbox requires the state space model of the system \eqref{eq:model}, its constraints \eqref{eq:constraints}, the ingredients of the MPCT formulation ($Q$, $R$, $T$, $S$ and $N$), and the parameters of the EADMM algorithm ($\rho$ and $\epsilon$).
It then generates library-free plain C code containing the sparse solver for its direct implementation in the embedded system, which we compile using the \textit{gcc} compiler in the Raspberry Pi.

To test the performance of the proposed MPCT solver we conduct two experiments on the real system: disturbance rejection and reference tracking.

\begin{figure*}[t]
    \centering
    \begin{subfigure}[ht]{\figwidth\textwidth}
        \includegraphics[width=\linewidth]{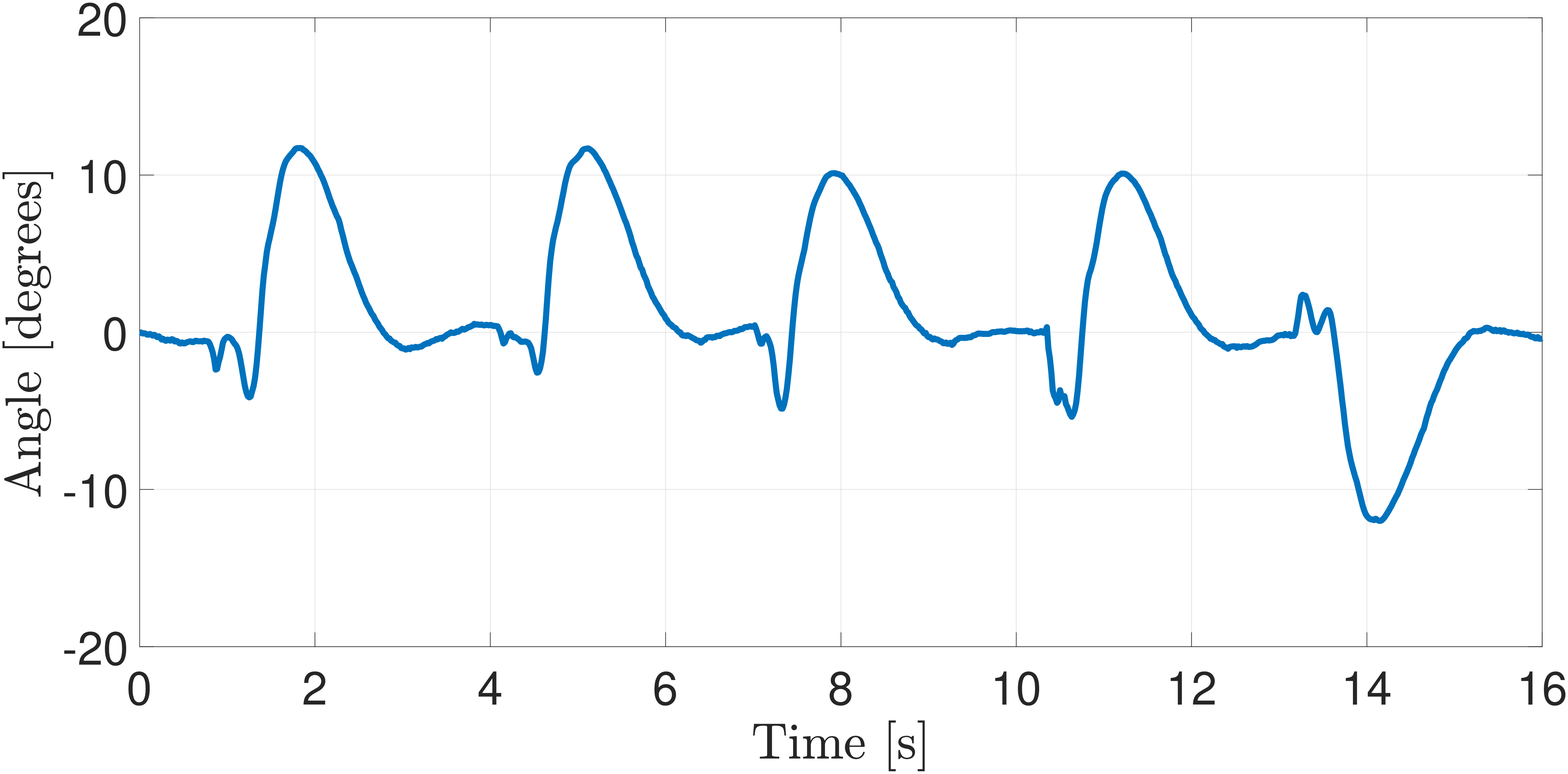}
        \caption{Tilte of the system.}
        \label{fig:pulse:angle}
    \end{subfigure}%
    \quad%%
    \begin{subfigure}[ht]{\figwidth\textwidth}
        \includegraphics[width=\linewidth]{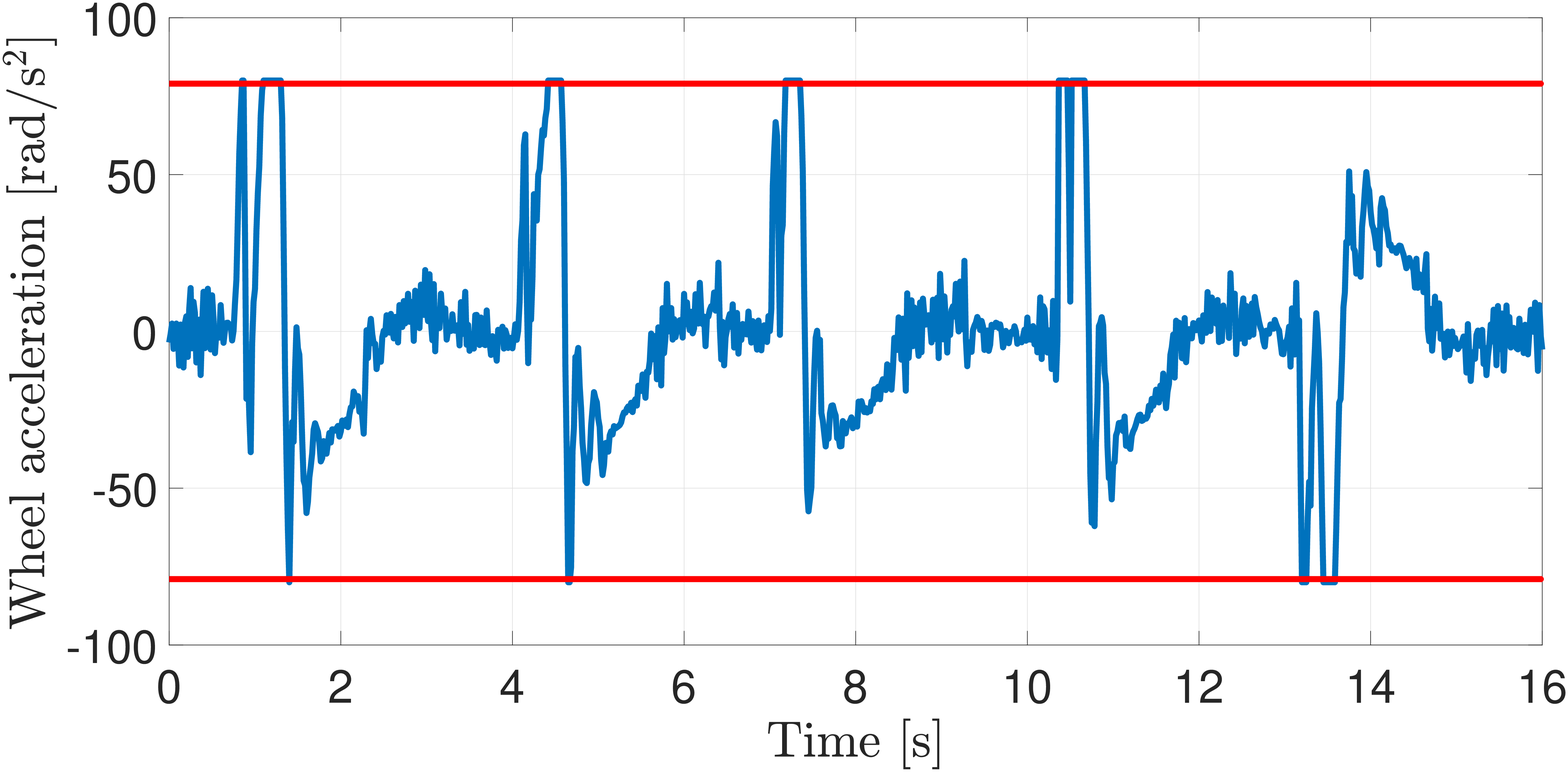}
        \caption{Angular acceleration of the wheels.}
        \label{fig:pulse:acceleration}
    \end{subfigure}%
    
    \begin{subfigure}[ht]{\figwidth\textwidth}
        \includegraphics[width=\linewidth]{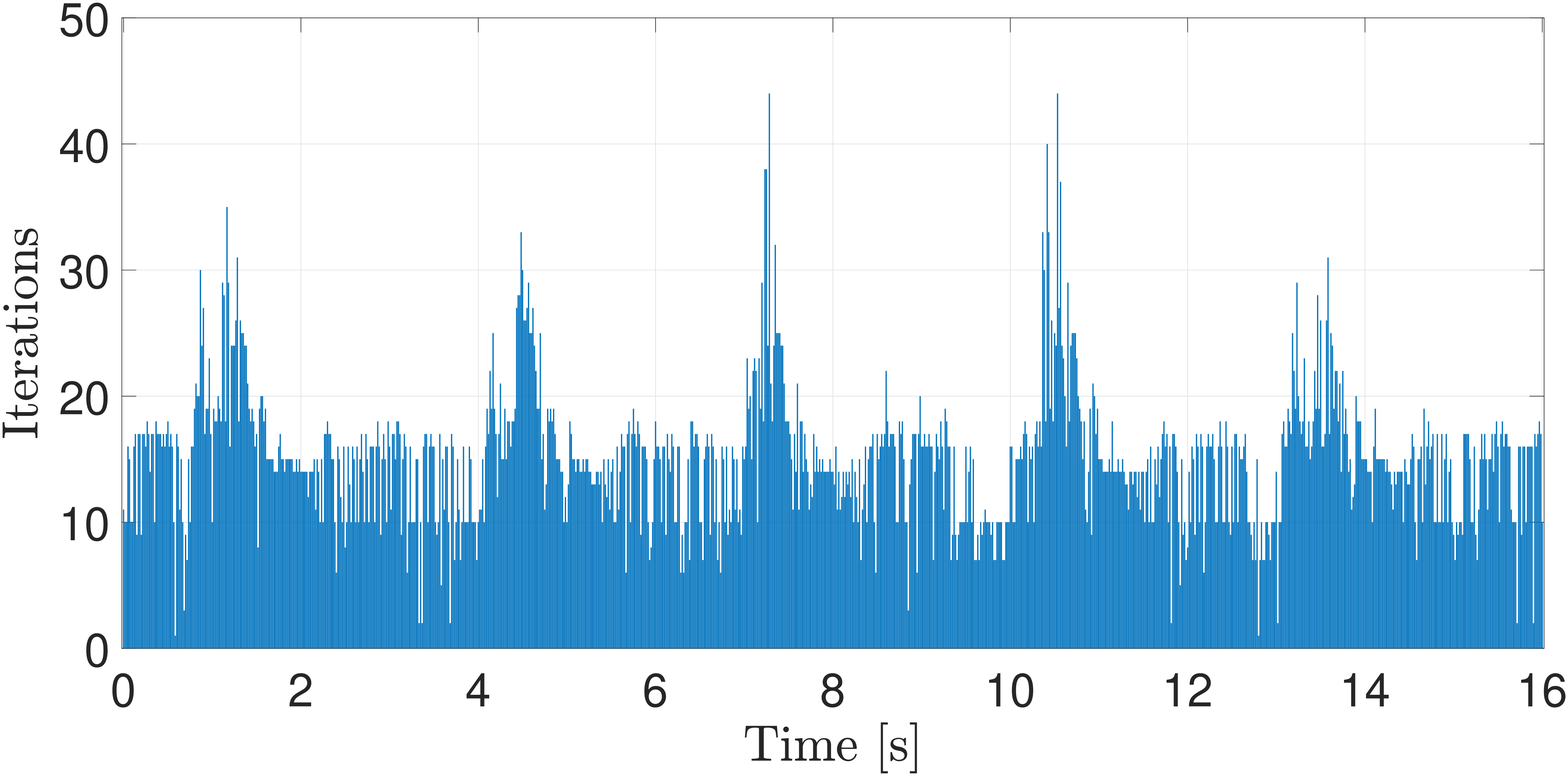}
        \caption{Iterations of the MPCT solver.}
        \label{fig:pulse:iterations}
    \end{subfigure}%
    \quad%%
    \begin{subfigure}[ht]{\figwidth\textwidth}
        \includegraphics[width=\linewidth]{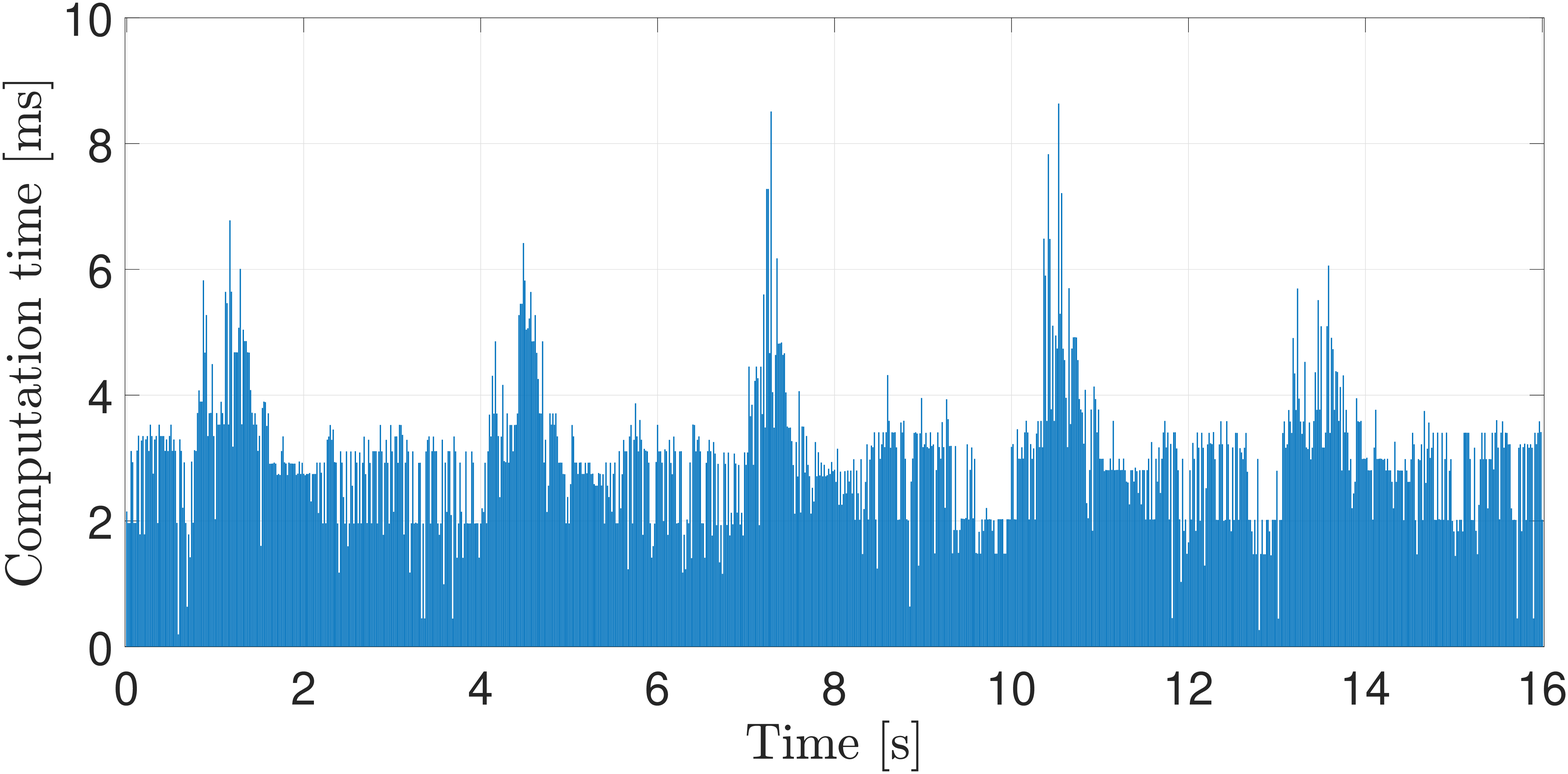}
        \caption{Computation times of the MPCT solver.}
        \label{fig:pulse:computation}
    \end{subfigure}%
    \caption{Closed-loop results of the robot against external disturbances.}
    \label{fig:pulse}
\end{figure*}

\begin{figure*}[t]
    \centering
    \begin{subfigure}[ht]{\figwidth\textwidth}
        \includegraphics[width=\linewidth]{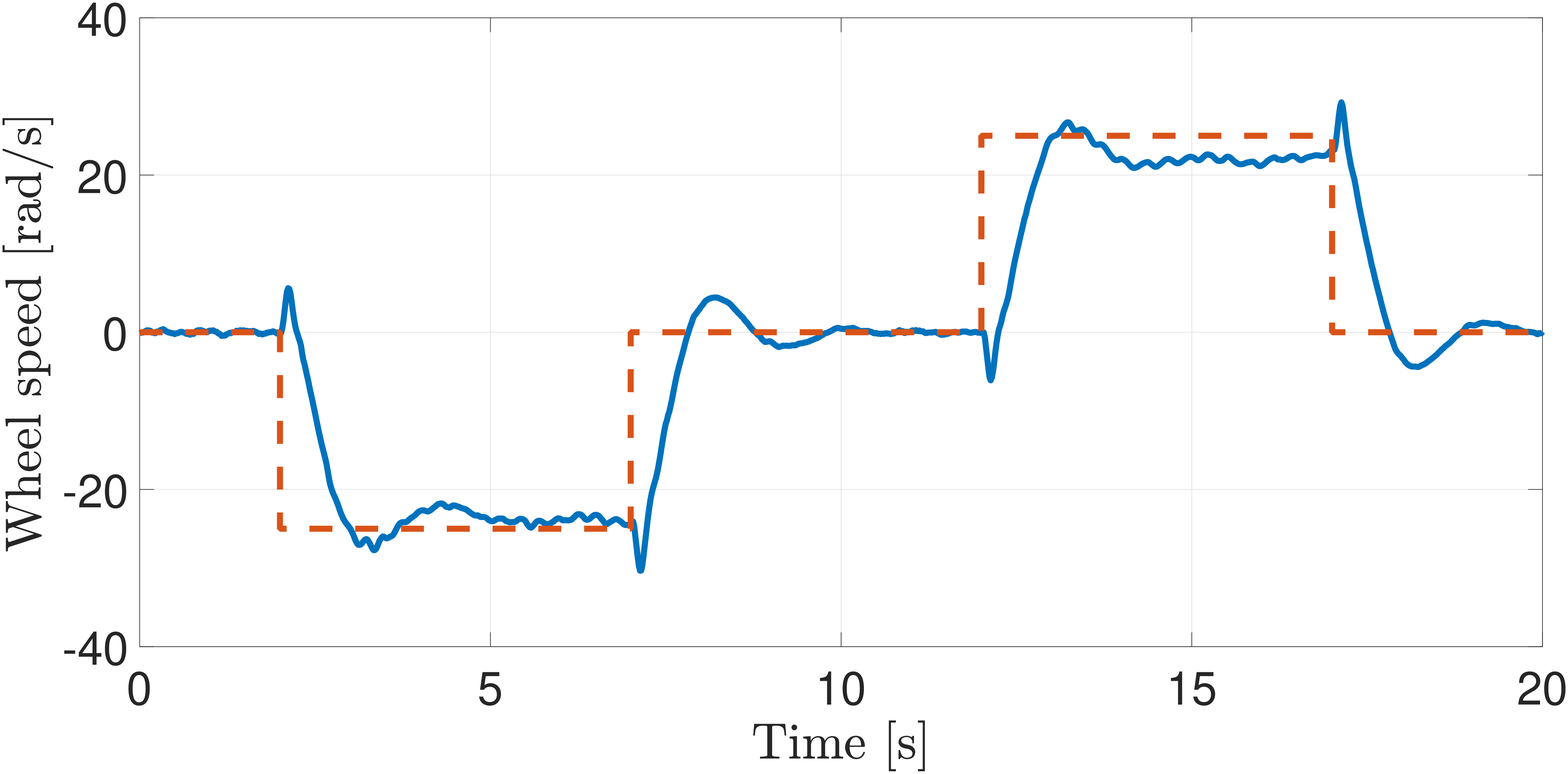}
        \caption{Angular speed of the wheels.}
        \label{fig:ref:speed}
    \end{subfigure}%
    \quad%%
    \begin{subfigure}[ht]{\figwidth\textwidth}
        \includegraphics[width=\linewidth]{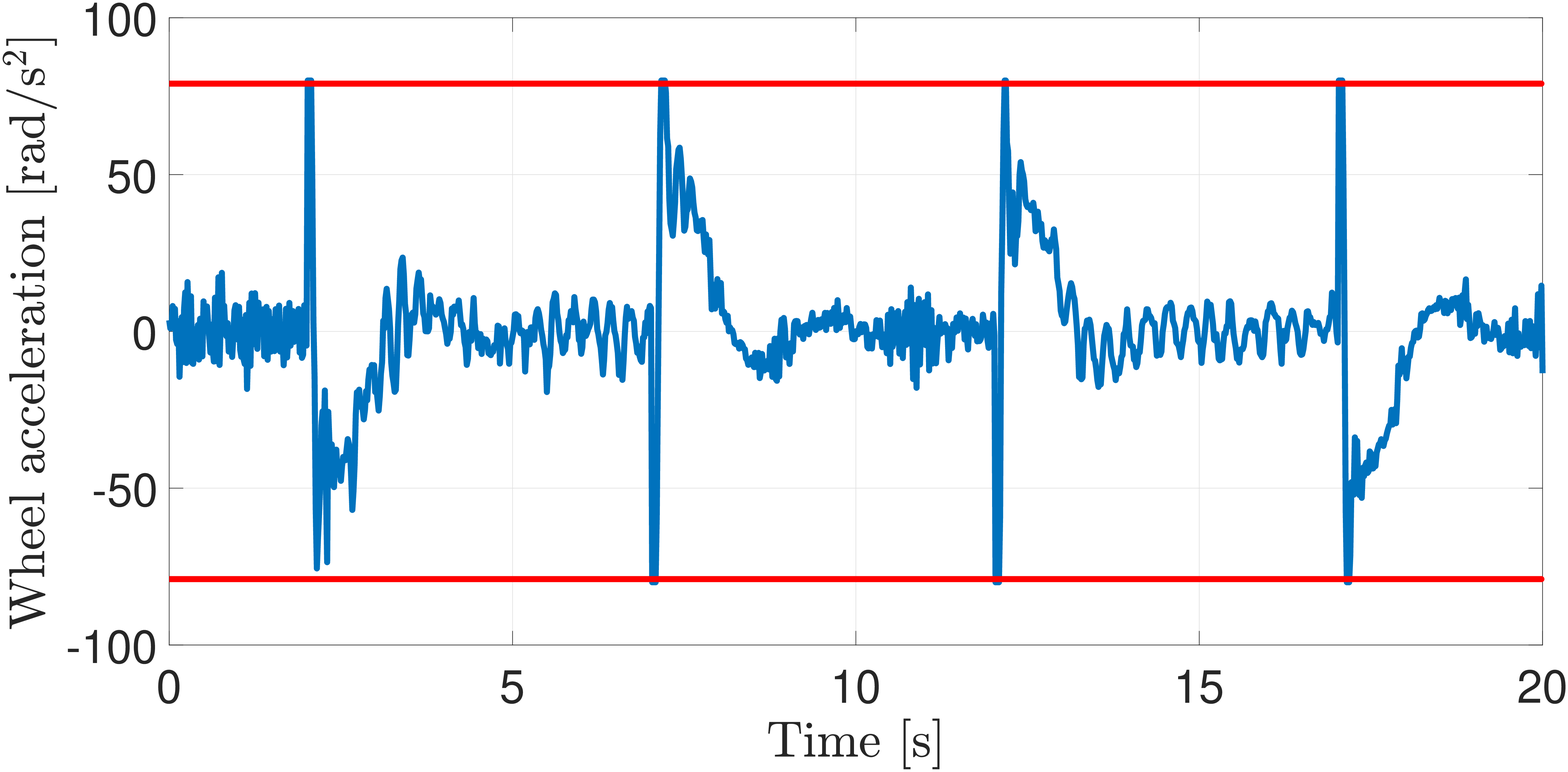}
        \caption{Angular acceleration of the wheels.}
        \label{fig:ref:acceleration}
    \end{subfigure}%
    
    \begin{subfigure}[ht]{\figwidth\textwidth}
        \includegraphics[width=\linewidth]{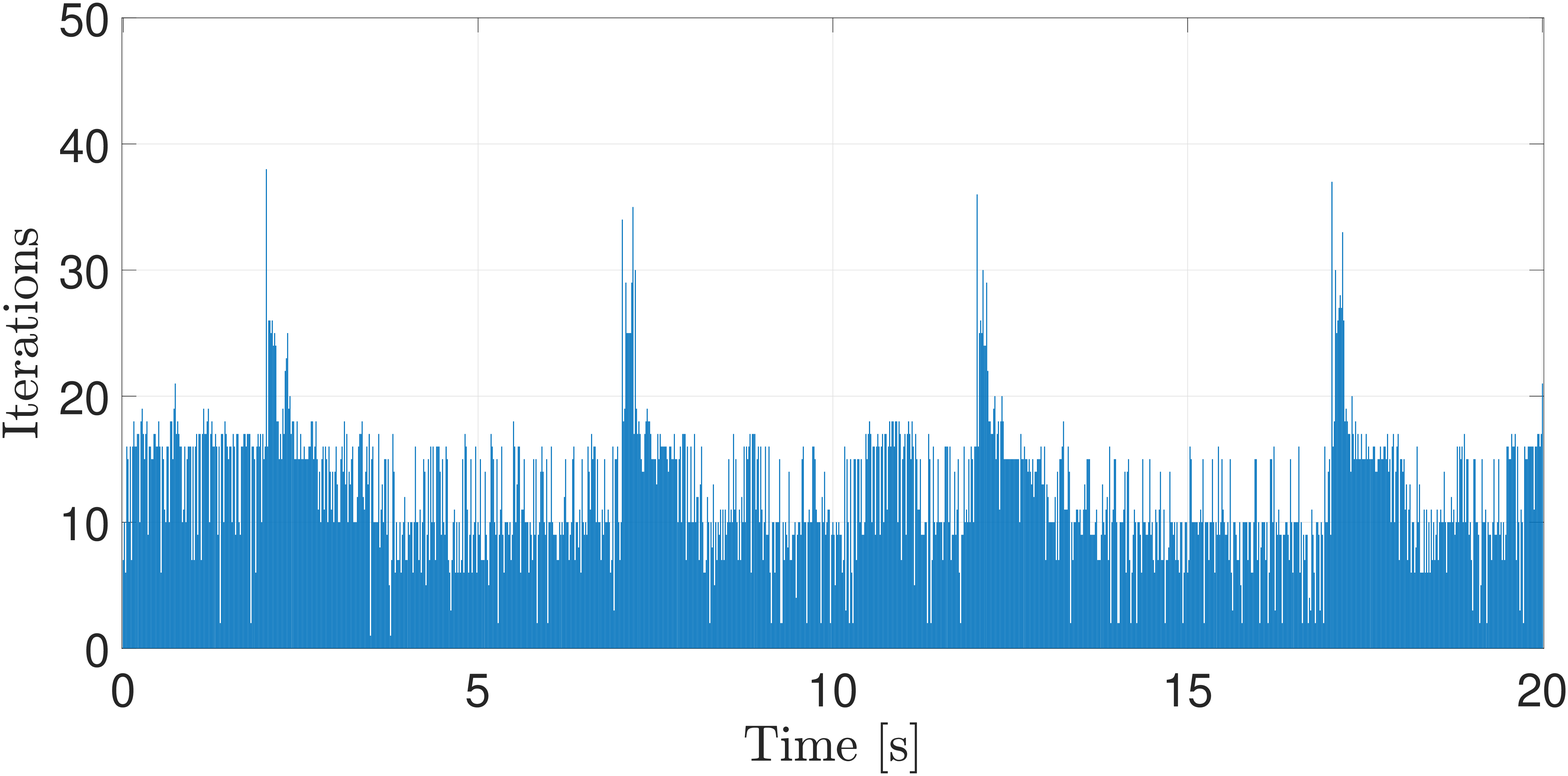}
        \caption{Iterations of the MPCT solver.}
        \label{fig:ref:iterations}
    \end{subfigure}%
    \quad%%
    \begin{subfigure}[ht]{\figwidth\textwidth}
        \includegraphics[width=\linewidth]{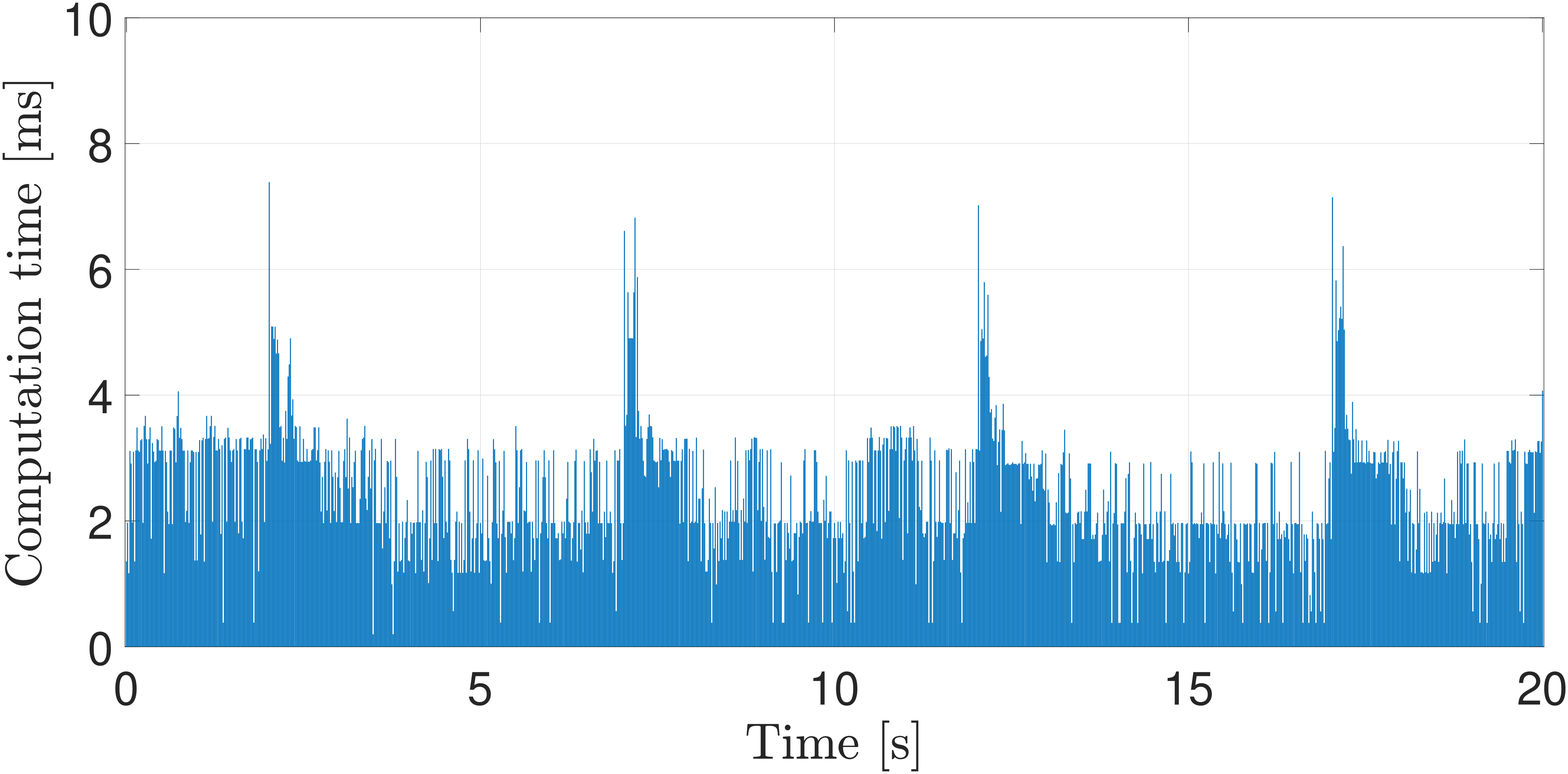}
        \caption{Computation times of the MPCT solver.}
        \label{fig:ref:computation}
    \end{subfigure}%
    \caption{Closed-loop results of the robot for changing reference of the angular speed.}
    \label{fig:ref}
\end{figure*}

Figure \ref{fig:pulse} shows the disturbance rejection results. In this experiment, the reference is set to the operating point, i.e., $x_r = x^\circ$ and $u_r = u^\circ$. The system is initially positioned at the reference and is then repeatedly perturbed by manually pushing it in either direction.
Figures \ref{fig:pulse:angle} and \ref{fig:pulse:acceleration} show the tilt of the system $\phi$ (in degrees), and the control action $\ddot \theta$, respectively. Figures \ref{fig:pulse:iterations} and \ref{fig:pulse:computation} show the number of iterations and computation time of the EADMM algorithm at each sample time, respectively.
As can be seen, the MPCT controller steers the system back to the vertical position after each push. Additionally, the control action reaches its upper and lower bounds, which are marked in red lines, during the first moments after each disturbance is applied. Note that the number of iterations of the solver increases when the control action bounds are active, as expected when using first-order methods. However, the increase is not very significant.

Figure \ref{fig:ref} shows the reference tracking results. In this experiment, the system is started at the operating point and then the reference for the wheel angular speed $\dot \theta$ is changed in multiple occasions.
Figures \ref{fig:ref:speed} and \ref{fig:ref:acceleration} show the speed of the wheels $\dot \theta$, and the control action $\ddot \theta$, respectively. Figures \ref{fig:ref:iterations} and \ref{fig:ref:computation} show the number of iterations and the computation time of the EADMM algorithm at each sample time, respectively.
As can be seen, the MPCT controller steers the system to the reference.
A slight offset can be observed for references other than the operating point due to the difference between the prediction model \eqref{eq:model} and the real system.
This offset could be corrected with the inclusion of a state and disturbance estimator \cite{Krupa_TCST_20}.
Once again, the control action reaches its upper and lower bounds during the first moments after each reference change, without having a significant impact on the number of iterations.

Table \ref{tab:case:study:computation:performance} shows a detailed analysis of the number of iterations and computation times of the algorithm during the two experiments. We show the maximum, minimum, median and average number of iterations and computation times.

\section{Conclusions} \label{sec:conclusions}

This paper presents the results of implementing the sparse solver for the MPC for tracking formulation presented in \cite{Krupa_arxiv_MPCT_20} in a Raspberry Pi to control an inverted pendulum robot with fast dynamics.
The proposed solver is available in the Spcies toolbox \cite{Spcies} for Matlab at \url{https://github.com/GepocUS/Spcies}.

The results indicate that the solver, which is based on an extension of the ADMM algorithm, is suitable for its implementation in embedded systems to control systems with sample times in the order of milliseconds in real-time.

% Fakesection Bibliography
\bibliographystyle{IEEEtran}
\bibliography{IEEEabrv, BibSegway}

\end{document}